# MESURE Tool to benchmark Java Card platforms

Samia Bouzefrane[1], Julien Cordry[1] and Pierre Paradinas[2]

[1]CEDRIC Laboratory, Conservatoire National des Arts et Métiers
292 rue Saint Martin, 75141, Paris Cédex 03, FRANCE
{last_name.first_name@cnam.fr}

[2]INRIA, Domaine de Voluceau - Rocquencourt -B.P. 105, 78153
Le Chesnay Cedex, FRANCE.
{ pierre.paradinas @inria.fr}

**Abstract**
The advent of the Java Card standard has been a major turning point in smart card technology. With the growing acceptance of this standard, understanding the performance behavior of these platforms is becoming crucial. To meet this need, we present in this paper a novel benchmarking framework to test and evaluate the performance of Java Card platforms. MESURE tool is the first framework which accuracy and effectiveness are independent from the particular Java Card platform tested and CAD used.

## 1. Introduction

With more than 5 billion copies in 2008 [2], smart cards are an important device of today's information society. The development of the Java Card standard made this device even more popular as it provides a secure, vendor-independent, ubiquitous Java platforms for smart cards. It shortens the time-to-market and enables programmers to develop smart card applications for a wide variety of vendors products. In this context, understanding the performance behavior of Java Card platforms is important to the Java Card community (users, smart card manufacturers, card software providers, card users, card integrators, etc.). Currently, there is no solution on the market which makes it possible to evaluate the performance of a smart card that implements Java Card technology. In fact, the programs which realize this type of evaluations are generally proprietary and not available to the whole of the Java Card community. Hence, the only existing and published benchmarks are used within research laboratories (e.g., SCCB project from CEDRIC laboratory [5] or IBM Research [12]). However, benchmarks are important in the smart card area because they contribute in discriminating companies products, especially when the products are standardized. In this paper, on one hand we propose a general benchmarking solution through different steps that are essential for measuring the performance of the Java Card platforms; on the other hand we validate the obtained measurements from statistical and precision CAD (Card Acceptance Device) points of view.

The remainder of this paper is organised as follows. In Section 2, we describe briefly some benchmarking attempts in the smart card area. In Section 3, an overview of the benchmarking framework is given. Section 4 analyses the obtained measurements using first a statistical approach, and then a precision reader, before concluding the paper in Section 5.

## 2. Java-Card Benchmarking State of the Art

Currently, there is no standard benchmark suite which can be used to demonstrate the use of the Java Card Virtual Machine (JCVM) and to provide metrics for comparing Java Card platforms. In fact, even if numerous benchmarks have been developed around the Java Virtual Machine (JVM), there are few works that attempt to evaluate the performance of smart cards. The first interesting initiative has been done by Castellà et al. in [4] where they study the performance of micro-payment for Java Card platforms, i.e., without PKI (Public Key Infrastructure). Even if they consider Java Card platforms from distinct manufacturers, their tests are not complete as they involve mainly computing some hash functions on a given input, including the I/O operations. A more recent and complete work has been undertaken by Erdmann in [6]. This work mentions different application domains, and makes the distinction between I/O, cryptographic functions, JCRE (Java Card Run Time Execution) and energy consumption. Infineon Technologies is the only provider of the tested cards for the different application domains, and the software itself is not available. The work of Fischer in [7] compares the performance results given by a Java Card applet with the results of the equivalent native application. Another interesting work has been carried out by the IBM BlueZ secure systems group and it was detailed in a Master thesis [12]. JCOP framework has been used to perform a series of tests to cover the communication overhead, DES performance and reading and writing operations into the card memory (RAM and EEPROM). Markantonakis in [9] presents some performance comparisons between the two most widely used terminal APIs, namely PC/SC and OCF. Comparatively to these works, our benchmarking framework not only covers the different functionalities of a Java Card platform but it also provided as a set of open source code freely accessible on-line.

## 3. General benchmarking framework

3.1 Introduction

Our research work falls under the MESURE project [10], a project funded by the French administration (Agence Nationale de Recherche), which aims at developing a set of open source tools to measure the performance of Java Card platforms. These benchmarking tools focus on Java Card 2.2 functionalities even if Java Card 3.0 specifications have been published since March 2008 [1], principally because until now there is no Java Card 3.0 platform in the market except for some prototypes such as the one demonstrated by Gemalto during the Java One Conference in June 2008. Since Java Card 3.0 proposes two editions: connected (web oriented) edition and classic edition, our measuring tools can be reused to benchmark Java Card 3.0 classic edition platforms.

3.2 Addressed issues

Only features related to the normal use phase of Java Card applications will be considered here. Excluded features include installing, personalizing or deleting an application since they are of lesser importance from user's point of view and performed once.

Hence, the benchmark framework enables performance evaluation at three levels:

– The VM level: to measure the execution time of the various instructions of the virtual machine (basic instructions), as well as subjacent mechanisms of the virtual machine (e.g., reading and writing the memory).

– The API level: to evaluate the functioning of the services proposed by the libraries available in the embedded system (various methods of the API, namely those of Java Card and GlobalPlatform).

– The JCRE (Java Card Runtime Execution) level: to evaluate the non-functional services, such as the transaction management, the method invocation in the applets, etc.

We will not take care of features like the I/Os or the power consumption because their measurability raises some problems such as:

– For a given smart card, distinct card readers may provide different I/Os measurements.

– Each part of an APDU is managed differently on a smart card reader. The 5 bytes header is read first, and the following data can be transmitted in several way: 1 acknowledge for each byte or not, delay or not before noticing the status word, etc.

– The smart card driver used by the workstation generally induces more delay on the measurement than the smart card reader itself.

3.3 The benchmarking overview

The set of tests are supplied to benchmark Java Card platforms available for anybody and supported by any card reader. The various tests thus have to return accurate results, even if they are not executed on precision readers. We reach this goal by removing the potential card reader weakness (in terms of delay, variance and predictability) and by controlling the noise generated by measurement equipment (the card reader and the workstation). Removing the noise added to a specific measurement can be done with the computation of an average value extracted from multiple samples. As a consequence, it is important on the one hand to perform each test several times and to use basic statistical calculations to filter the trustworthy results. On the other hand, it is necessary to execute several times in each test the operation to be measured in order to fix a minimal duration for the tests (> 1 second) and to expect getting precise results. We defined a set of modules as part of the benchmarking framework. The benchmarks have been developed under the Eclipse environment based on JDK 1.6, with JSR268 [13] that extends Java Standard Edition with a package that defines methods within Java classes to interact with a smart card. According to the ISO 7816 standard, since a smart card has no internal clock, we are obliged to measure the time a Java Card platform takes to answer to an APDU command, and to use that measure to deduce the execution time of some operations.

The benchmarking development tool covers two parts as described in Figure 1: the script part and the applet part. The script part, entirely written in Java, defines an abstract class that is used as a template to derive test cases characterized by relevant measuring parameters such as, the operation type to measure, the number of loops, etc. A method *run*() is executed in each script to interact with the corresponding test case within the applet. Similarly, on the card is defined an abstract class that defines three methods:
– a method *setUp*() to perform any memory allocation needed during the lifetime test case.
– a method *run*() used to launch the tests corresponding to the test case of interest, and
– a method *cleanUp*() used after the test is done to perform any clean-up.

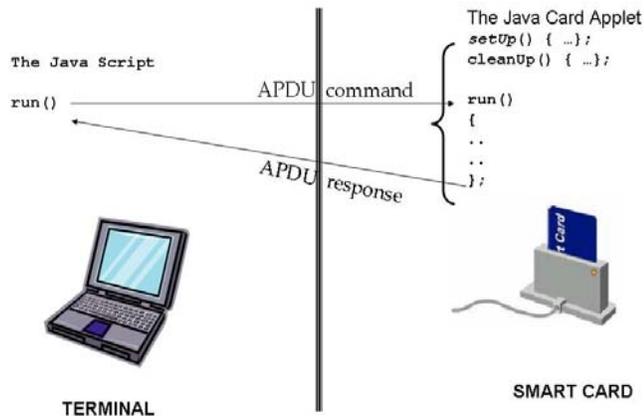

Fig. 1 The script part and the Applet part

## 3.4 Modules

In this section, we describe the general benchmark framework (see Figure 2) that has been designed to achieve the MESURE goal. The methodology consists of different steps. The objective of the first step is to find the optimal parameters used to carry out correctly the tests. The tests cover the Virtual Machine (VM) operations and the API methods. The obtained results are filtered by eliminating non-relevant measurements and values are isolated by drawing aside measurement noise. A profiler module is used to assign a mark to each benchmark type, hence allowing us to establish a performance index for each smart card profile used. In the following subsections, we detail every module composing the framework.

The bulk of the benchmark consists in performing time execution measurements when we send APDU commands from the computer through the CAD to the card. Each test (through the `run` method) is performed within the card a certain number of times (Y) to ensure reliability of the collected execution times, and within each `run` method, we perform a certain number of loops (L). L is coded on the byte P2 of the APDU commands which are sent to the on-card applications. The size of the loop performed on the card is $L = (P2)^2$ since L is so great to be represented with one byte.

**The Calibrate Module:** computes the optimal parameters (such as the number of loops) needed to obtain measurements of a given precision.

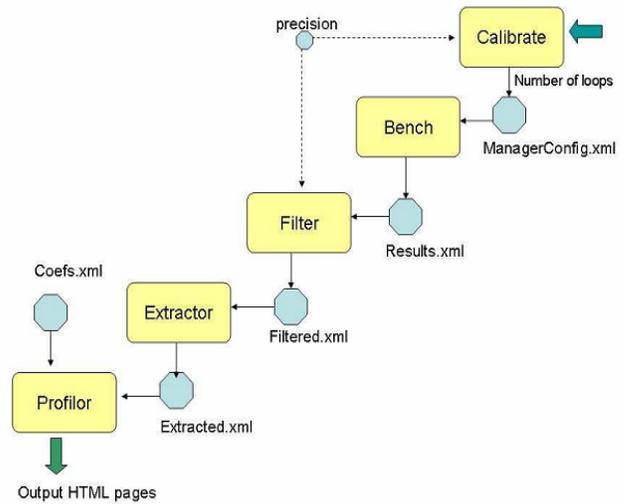

Fig. 2 Overall Architecture

Benchmarking the various different byte-codes and API entries takes time. At the same time, it is necessary to be precise enough when it comes to measuring those execution times. Furthermore, the end user of such a benchmark should be allowed to focus on a few key elements with a higher degree of precision. It is therefore necessary to devise a tool that let us decide what are the most appropriate parameters for the measurement.

Figure 3 depicts the evolution of the raw measurement, as well as its standard deviation, as we take 30 measurements for each available loop size of a test applet. As we can see, the measured execution time of an applet grows linearly with the number of loops being performed on the card (L). On the other hand, the perceived standard deviation on the different measurements varies randomly as the loop size increases, though with less and less peaks. Since a bigger loop size means a relatively more stable standard deviation, we use both the standard deviation and the mean measured execution time as a basis to assess the precision of the measurement.

To assess the reliability of the measurements, we compare the value of the measurement with the standard deviation. The end user will need to specify this ratio between the average measurement and the standard deviation, as well as an optional minimum accepted value, which is set at one second by default. The ratio refers to the precision of the tests while the minimal accepted value is the minimum duration to perform each test. Hence, with both the ratio and the minimal accepted value, as specified by the end user, we can test and try different values for the loop size to binary search and approach the ideal value.

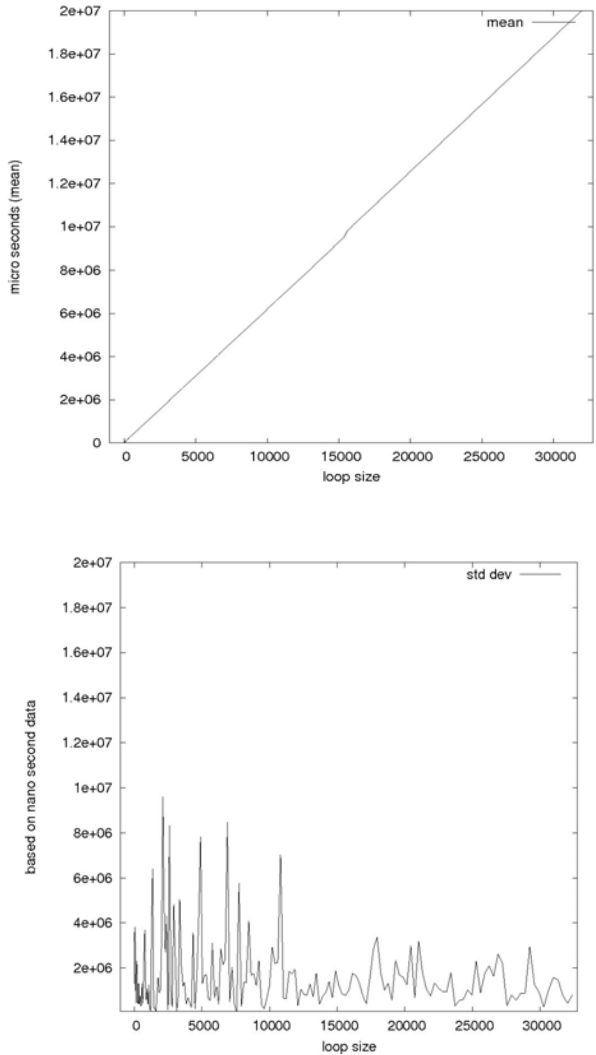

Fig. 3 Raw measurements and Standard deviation

**The Bench Module:** For a number of cycles, defined by the calibrate module, the bench module computes the execution time for:
- The VM byte codes
- The API methods
- The JCRE mechanisms (such as transactions).

**The Filter Module:** Experimental errors lead to noise in the raw measurement experiments. This noise leads to imprecision in the measured values, making it difficult to interpret the results. In the smart card context, the noise is due to crossing the platform, the CAD and the terminal (measurement tools, Operating System, hardware).

The issues become: how to interpret the varying values and how to compare platforms when there is some noise in the results. The filter module uses a statistical design to extract meaningful information from noisy data. From multiple measurements for a given operation, the filter module uses the mean value µ of the set of measurements to guess the actual value, and the standard deviation σ of the measurements to quantify the spread of the measurements around the mean. Moreover, since the measurements respect the normal Gaussian distribution, a confidence interval $[\mu - (n \times \sigma), \mu + (n \times \sigma)]$, within which the confidence level is of $1-a$, is used to help eliminate the measurements outside the confidence interval, where $n$ and $a$ are respectively the number of measurements and the temporal precision, and they are related by traditional statistical laws.

**The Extractor Module:** is used to isolate the execution time of the features of interest among the mass of raw measurements that we gathered so far. Benchmarking byte-codes and API methods within Java Card platforms requires some subtle means in order to obtain execution results that reflect as accurately as possible the actual isolated execution time of the feature of interest. This is because there exists a significant and non-predictable elapse of time between the beginning of the measure, characterized by the starting of the timer on the computer, and the actual execution of the byte-code of interest. This is also the case the other way around. Indeed, when performing a request on the card, the execution call has to travel several software and hardware layers down to the card's hardware and up to the card's VM (vice versa upon response). This non-predictability is mainly dependent on hardware characteristics of the benchmark environment (such as the CAD, PC's hardware, etc), the Operating System level interferences, services and also on the PC's VM.

To minimize the effect of these interferences, we need to isolate the execution time of the features of interest, while ensuring that their execution time is sufficiently important to be measurable. The maximization of the byte-codes execution time requires a test applet structure with a loop having a large upper bound, which will execute the byte-codes for a substantial amount of time. On the other hand, to achieve execution time isolation, we need to compute the isolated execution time of any auxiliary byte-code upon which the byte-code of interest is dependent. For example if *sadd* is the byte-code of interest, then the byte-codes that need to be executed prior to its execution are those in charge of loading its operands onto the stack, like two *sspush*. Thereafter we subtract the execution time of an empty loop and the execution time of the auxiliary byte-codes from that of the byte-code of interest ($op_n$ in Table 1) to obtain the isolated execution time of the byte-code. As presented in Table 1, the actual test is performed within a method `run` to ensure that the stack is freed after each invocation, thus guaranteeing memory availability.

Table 1: The framework for a bytecode $op_n$

| *Java Card Applet* | *Test Case* |
|---|---|
| **process**() { <br>   i = 0 <br>   while i <= L <br>     do { <br>       run() <br>       i = i+1 <br>     } <br> } | **run**() { <br>   $op_0$ <br>   $op_1$ <br>   . <br>   . <br>   . <br>   $op_{n-1}$ <br>   **$op_n$** <br> } |

In Table 1:
- *L* represents the chosen upper bound,
- $op_n$ represents the byte-code of interest,
- $op_i$ for i ∈ [0..n-1] represents the auxiliary byte-codes necessary to perform the byte-code $op_n$.

To compute the mean isolated execution time of $op_n$ we need to perform the following calculation:

$$\overline{M(op_n)} = \frac{\overline{m_L(op_n)} - \overline{m_L(Emptyloop)}}{L} - \sum_{i=0}^{n-1} \overline{M(op_i)} \quad (1)$$

Where :

- $\overline{M(op_i)}$ is the mean isolated execution time of the byte-code $op_i$
- $\overline{m_L(op_n)}$ is the mean global execution time of the byte-code $op_n$, including interferences coming from other operations performed during the measurement, both on the card and on the computer, with respect to a loop size *L*. These other operations represent for example auxiliary byte-codes needed to execute the byte-code of interest, or OS and JVM specific operations. The mean is computed over a significant number of tests. It is the only value that is experimentally measured.
- *Emptyloop* represents the execution of a case where the `run` method does nothing.

The formula (1) implies that prior to computing $\overline{M(op_n)}$ we need to compute $\overline{M(op_i)}$ for i ∈ [0..n-1].

**The Profiler Module:** In order to define performance references, our framework provides measurements that are specifically adapted to one of the following application domains:

– Banking applications
– Transport applications, and
– Identity applications.

A JCVM is instrumented in order to count the different operations performed during the execution of a script for a given application. More precisely, this virtual machine is a simulated and proprietary VM executing on a workstation. This instrumentation method is rather simple to implement compared to a static analysis based methods, and can reach a good level of precision, but it requires a detailed knowledge of the applications and of the most significant scripts.

Some features related to byte-codes and API methods appeared to be necessary and the simulator was instrumented to give useful information such as:
– for the API methods :
  • the types and values of method parameters
  • the length of arrays passed as parameters,
– for the byte-codes :
  • the type and duration of arrays for array related byte-codes (*load*, *astore*, *arraylength*),
  • the transaction status when invoking the byte-code.

A simple utility tool has been developed to parse the log files generated by the instrumented JCVM, which builds a human-readable tree of method invocations and byte-code usage. Thus, with the data obtained from the instrumented VM, we attribute for each application domain a number that represents the performance of some representative applets of the domain on the tested card. Each of these numbers is then used to compute a global performance mark. We use weighted means for each domain dependent mark. Those weights are computed by monitoring how much each Java Card feature is used within a regular use of standard applets for a given domain. For instance, if we want to test the card for a use in transport applications, we will use the statistics that we gathered with a set of representative transport applets to evaluate the impact of each feature of the card.

We are considering the measure of the feature *f* on a card *c* for an application domain *d*. For a set of $n_M$ extracted measurements $M^1_{c,f}$, …, $M^{nM}_{c,f}$ considered as significant for the feature *f*, we can determine a mean $\overline{M_{c,f}}$ modelling the performance of the platform for this feature. Given $n_C$ cards for which the feature *f* was measured, it is necessary to determine the reference mean execution time $R_f$, which will then serve as a basis of comparison for all subsequent test. Hence the "mark" $N_{cf}$ of a card *c* for a feature *f*, is the relation between $R_f$ and $\overline{M_{c,f}}$ :

$$N_{c,f} = \frac{R_f}{\overline{M_{c,f}}} \quad (2)$$

However, this mark is not weighted. For each pair of a feature $f$ and an application domain $d$, we associate a coefficient $\alpha_{f,d}$, which models the importance of $f$ in $d$. The more a feature is used within typical applications of the domain, the bigger the coefficient:

$$\alpha_{f,d} = \frac{\beta_{f,d}}{\sum_{i=1}^{n_F}\beta_{i,d}} \quad (3)$$

where :
– $\beta_{f,d}$ is the total number of occurrence of the feature $f$ in typical applications of the domain $d$.
– $n_F$ is the total number of features involved in the test.

Therefore, the coefficient $\alpha_{f,d}$ represents the occurrence proportion of the feature of interest $f$ among all the features.

Hence, given a feature $f$, a card $c$ and a domain $d$, the "weighted mark" $W_{c,f,d}$ is computed as follows :
$$W_{c,f,d} = N_{c,f} \times \alpha_{f,d} \quad (4)$$

The "global mark" $P_{c,d}$ for a card $c$ and for a domain $d$ is then the sum of all weighted marks for the card. A general domain independent note for a card is computed as the mean of all the domain dependant marks.

Figure 4 shows some significant byte-codes computed for a card and compared to the reference tests regarding the financial domain. Whereas, Figure 5 shows the global results obtained for a tested card. Based on the results of Figure 5, our tested card seems to be dedicated for financial use.

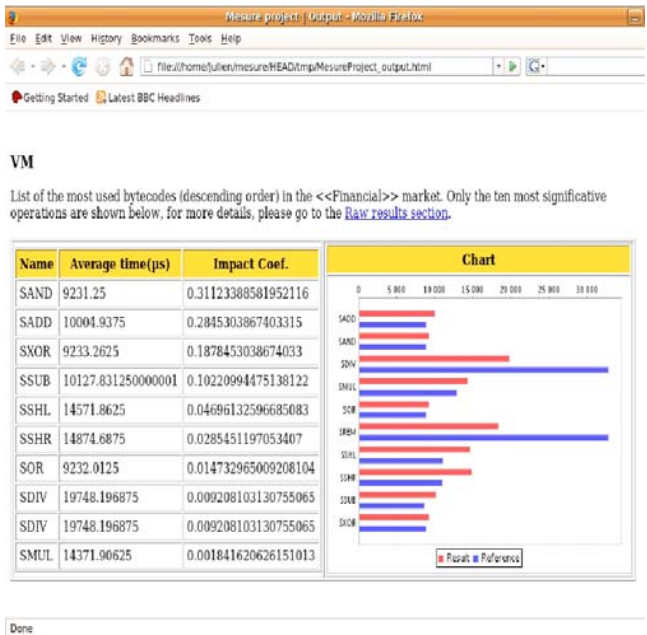
Fig. 4 An example of a financial-dependent mark

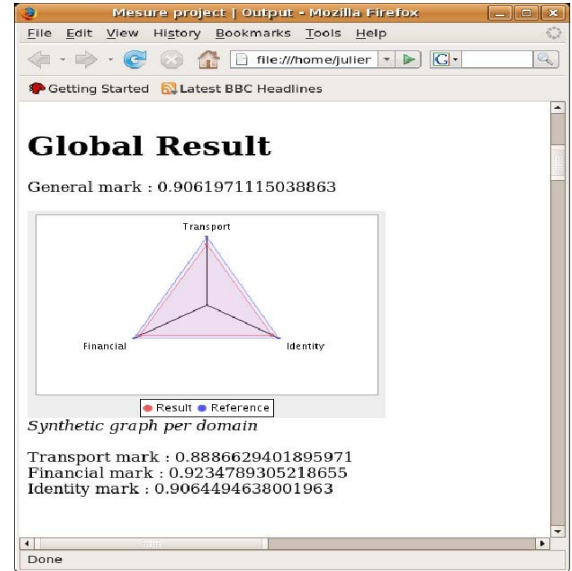
Fig. 5 Computing a global performance mark

## 4. Validation of the tests

### 4.1 Statistical correctness of the measurements

The expected distribution of any measurement is a normal distribution. The results being time values, if the distribution is normal, then, according to Lilja [8], the arithmetic mean is an acceptable representative time value for a certain number of measurements (Lilja recommends at least 30 measurements). Nevertheless, Rehioui [12] pointed out that the results obtained via methods similar to ours were not normally distributed on IBM JCOP41 cards. Erdmann [6] cited similar problems with Infineon smart cards. When we measure both the reference test and the operation test on several smart cards by different providers using different CADs on different OSs, none of the time performances had a normal distribution (see Figure 6 for a sample reference test performed on a card). The results were similar from one card to another in terms of distribution, even for different time values, and for different loop sizes. Changes in CAD, in host-side JVM, in task priority made no difference on the experimental distribution curve. Testing the cards on Linux and on Windows XP or Windows Vista, on the other side, showed differences. Indeed, the recurring factor when measuring the performances with a terminal running Linux with PC/SC Lite and a CCID driver is the gap between peaks of distribution. The peaks are often separated by 400ms and 100 ms steps which match some parts of the public code of PC/SC Lite and the CCID driver. With other CADs, the distribution shows similar steps with respect to the CAD driver source code. The peaks in the distribution from the measurements obtained on Windows are separated by 0.2 ms steps (see Figure 7). Without having access to neither the source code of the PC/SC implementation on Windows nor the driver source codes, we can deduce that there must be some similarities in the source codes between the proprietary versions and the open source versions.

In order to check the normality of the results, we isolated some of the peaks of some distribution obtained on our measurements and we used the resulting data set. The Shapiro-Wilk test is a well established statistical test used to verify the null hypothesis that a sample of data comes from a normally distributed population. The result of such a test is a number $W \in [0, 1]$, with W close to 1 when the data is normally distributed. No set of value obtained by isolating a peak within a distribution gave us a satisfying W close to 1. For instance, considering the peak in Figure 8, W = 0.8442, which is the highest value for W that we observed, with other values ranging as low as W = 0.1384. We conclude that the measurements we obtain, even if we consider a peak of distribution, are not normally distributed.

### 4.2 Validation through a precision CAD

We used a Micropross MP300 TC1 reader to verify the accuracy of our measurements. This is a smart card test platform, that is designed specifically to give accurate results, most particularly in terms of time analysis.

The results here are seemingly unaffected by noises on the host machine. With this test platform, we can precisely monitor the polarity changes on the contact of the smart card, that mark the I/Os.

We measured the time needed by a given smart card to reply to the same APDUs that we used with a regular CAD. We then tested the measured time values using the Shapiro-Wilk test, we observed $W \geq 0.96$, much closer to what we expected in the first place. So we can assume that the values are normally distributed for both the operation measurement and the reference measurement.

We subtracted each reference measurement value from each *sadd* operation measurement value, divided by the loop size to get a time values set that represents the time performance of an isolated *sadd* bytecode. Those new time values are normally distributed as well (W = 0.9522). On the resulting time value set, the arithmetic mean is 10611.57 ns and the standard deviation is 16.19524. According to [6], since we are dealing with a normal distribution, this arithmetic mean is an appropriate evaluation of the time needed to perform a *sadd* byte code on this smart card. Using a more traditional CAD (here, a Cardmann 4040, but we tried five different CADs) we performed 1000 measurements of the *sadd* operation test and 1000 measurements of the corresponding reference test. By subtracting each value obtained with the reference test from each of the values of the *sadd* operation test, and dividing by the loop size, we produced a new set of 1000000 time values. The new set of time values has an arithmetic mean of 10260.65 ns and a standard deviation of 52.46025.

The value we found with a regular CAD under Linux and without priority modification is just 3.42% away from the more accurate value found with the precision reader. Although this is a set of measurements that are not normally distributed (W = 0.2432), the arithmetic mean of our experimental noisy measurements seems to be a good approximation of the actual time it takes for this smart card to perform a *sadd*. The same test under Windows Vista gave us a mean time of 11380.83 ns with a standard deviation of 100.7473, that is 7,24% away from the accurate value.

We deduce that our data are noisy and faulty but despite a potentially very noisy test environment, our time measurements always provide a certain accuracy and a certain precision.

## 5. Conclusion

With the wide use of Java in smart card technology, there is a need to evaluate the performance and characteristics of these platforms in order to ascertain whether they fit the requirements of the different application domains. For the time being, there is no other open source benchmark solution for Java Card. The objective of our project [10] is to satisfy this need by providing a set of freely available tools, which, in the long term, will be used as a benchmark standard. In this paper, we have presented the overall benchmarking framework. Despite the noise, our framework achieves some degree of accuracy and precision. Our benchmarking framework does not need a costly reader to accurately evaluate the performance of a smart card. Java Card 3.0 is a new step forward for this community. Our framework should still be relevant to the classic edition of this platform, but we have yet to test it.

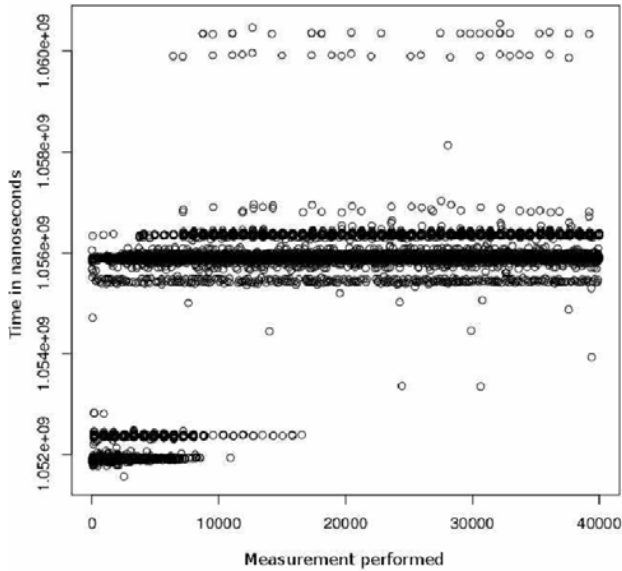

Fig. 6 Measurements of a reference test as the tests proceed under Linux, and the corresponding distribution curve L = $41^2$

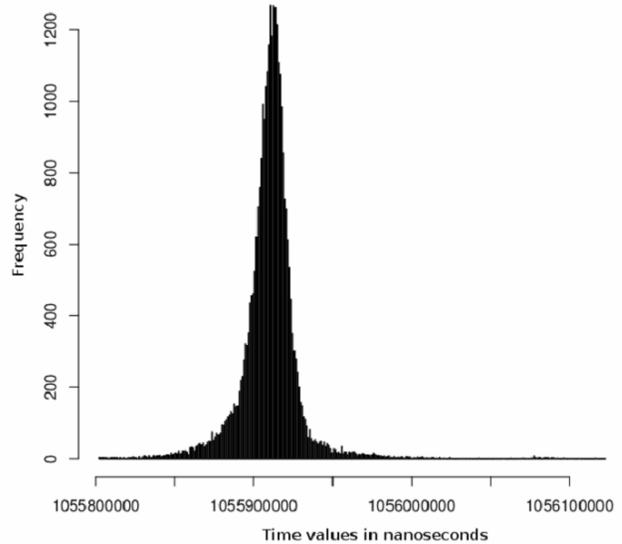

Fig. 8 Some Distribution of the measurement of a reference test: close up look at a peak in distribution L = $41^2$

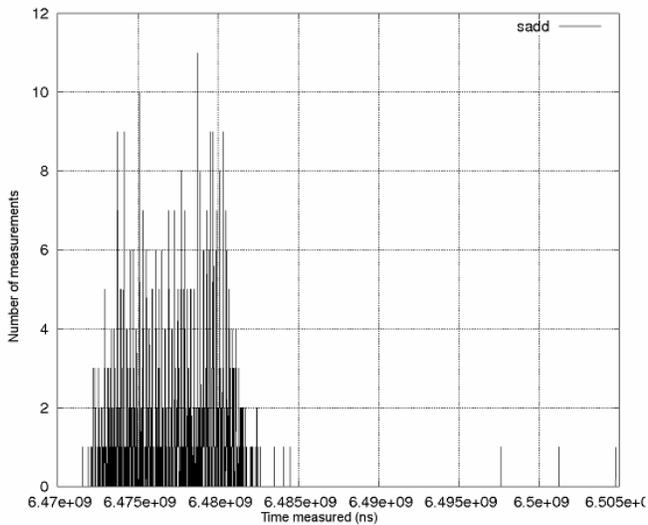

Fig. 7 Distribution of *sadd* operation measurements using Windows Vista, and a close up look at the distribution (L = $90^2$)

**Samia Bouzefrane** is an associate professor at the CNAM (Conservatoire National des Arts et Métiers) in Paris. She received her Ph. D. in Computer Science in 1998 at the University of Poitiers (France). She joined the CEDRIC Laboratory of CNAM on September 2002 after 4 years at the University of Le Havre. After many research works on real-time systems, she is interested in smart card area. Furthermore, she is the author of two books: a French/English/Berber dictionary (1996) and a book on operating systems (2003). Currently, she is a member of the ACM-SIGOPS, France Chapter.

**Julien Cordry** is a PhD student from the SEMpIA team (embedded and mobile systems towards ambient intelligence) of the CNAM in Paris. The topic of his research is the performance evaluation of Java Card platforms. He took part in the MESURE project, a collaborative work between the CNAM, the university of Lille and Trusted Labs. He gives lecturers at the CNAM, at the ECE (Ecole Centrale d'Electronique) and at the EPITA (a computer science engineering school). The MESURE project has received on September 2007 the Isabelle Attali Award from INRIA, which honors the most innovative work presented during "e-Smart" Conference.

**Pierre Paradinas** is currently the Technology-Development Director at INRIA, France. He is also Professor at CNAM (Paris) where he manages the "chair of Embedded and Mobile Systems". He received a PhD in Computer Science from the University of Lille (France) in 1988 on smart cards and health application. He joined Gemplus in 1989, and was successively researcher, internal technology audit, Advanced Product Manager while he launched the card based on Data Base engine (CQL), and the Director of a common research Lab with universities and National Research Center (RD2P). He sets up the Gemplus Software Research Lab in 1996. He was also appointed technology partnership Director in 2001 based in California until June 2003. He was the Gemplus representative at W3C, ISO/AFNOR, Open Card Framework and Java Community Process, co-editor of the part 7 of ISO7816, Director of the European funded Cascade project where the first 32-Risc microprocessor with Java Card was issued.